\documentclass[journal,10pt]{IEEEtran}

\usepackage{amsthm,amsmath,amssymb}

\usepackage[pdftex]{graphicx}
\graphicspath{{../pdf/}{../jpeg/}}
\DeclareGraphicsExtensions{.pdf,.jpeg,.png}

\begin{document}

\title{Deep Reinforcement Learning-based Anti-jamming Power Allocation in a Two-cell NOMA Network}

\author{Sina~Yousefzadeh~Marandy, Mohammad~Ali~Amirabadi,  Mohammad~Hossein~Kahaei,  and Seyed~Mohammad~Razavizadeh

\thanks{The authors are with the School of Electrical Engineering, Iran University
of Science and Technology, Tehran 16846-13114, Iran (e-mail:
sina\_yousefzadeh96@elec.iust.ac.ir; m\_amirabadi@elec.iust.ac.ir;  kahaei@iust.
ac.ir; smrazavi@iust.ac.ir).}
}

\markboth{
}
{}

\maketitle

\begin{abstract}
The performance of Non-orthogonal Multiple Access (NOMA) system dramatically decreases in the presence of inter-cell interference. This condition gets more challenging if a smart jammer is interacting in a network. In this paper, the NOMA power allocation of two independent Base Stations (BSs) against a smart jammer is, modeled as a sequential game. In this game, at first, each BS as a leader independently chooses its power allocation strategy. Then, the smart jammer as the follower selects its optimal strategy based on the strategies of the BSs. The solutions of this game are, derived under different conditions. Based on the game-theoretical analysis, three new schemes are proposed for anti-jamming NOMA power allocation in a two-cell scenario called a) Q-Learning based Unselfish (QLU) NOMA power allocation scheme, b) Deep Q-Learning based Unselfish (DQLU) NOMA power allocation scheme, and c) Hot Booting Deep Q-Learning based Unselfish (HBDQLU) NOMA power allocation scheme. In these methods the BSs do not coordinate with each other. But our analysis theoretically predicts that with high probability, the proposed methods will converge to the optimal strategy from the total network point of view. Simulation results show the convergence of the proposed schemes and also their outperformance with respect to the Q-Learning-based Selfish (QLS) NOMA power allocation method.
\end{abstract}


\IEEEpeerreviewmaketitle

\section{Introduction}\label{Introduction}

Power domain Non-Orthogonal Multiple Access (NOMA) schemes could considerably improve the network capacity and have superior spectral efficiency [1]-[5]. Using NOMA can also improve the outage performance and user fairness [6].
Inter-cell interference in multi-cell scenarios makes NOMA optimization problems more complicated [3]. Also, the  performance of NOMA networks; especially the performance of the cell-edge users, will significantly reduce due to the inter-cell interference [5]. When there is inter-cell interference in a network, the power allocation strategy in a single-cell scenario, will not work properly [7].

The wireless networks, especially the NOMA wireless networks, are vulnerable to smart jamming attacks [6]. A smart jammer by using programmable radio devices, such as software-defined radios, will be able to choose its jamming policy optimally and adaptively [6], [8], [9]. In NOMA networks the smart jammer can easily, lead the data rate of weaker users, to be less than QoS criteria.

When there is an inter-cell interference in a NOMA network, meanwhile, a smart jammer is interacting in the environment, it will be extremely challenging to meet QoS for all of the users. This problem is more sophisticated when there is not any coordination between Base Stations (BSs).

There are a lot of security issues in wireless networks which, traditional and system-based solutions could not solve them [10]. The authors of [10], has comprehensively studied, using the economic and pricing approaches, to deal with the security issues in wireless networks. Some references like [6] and [8] have studied the anti-jamming communication from a game theory point of view.

 Reinforcement Learning (RL) algorithms, like Q-Learning (QL), are widely used for different purposes in communication environments, such as the works in [6], [9], [11]. When a QL-based algorithm, works in a Markov decision process, it will find, the optimal policy. In some cases, due to the high dimensionality of states and actions space, multi-agents RL algorithm, have been designed, based on the coordination of the agents. In these schemes, they make their decision in order. In other words, they cannot interact with the environment, simultaneously. More than that, coordination is always costly and raises the complexity of the considering algorithm.

 There is a trend to use independent QL agents in a common environment. The interactions of multiple QL agents, in an environment, can be modeled as an evolutionary game in which, the agent will converge to the pure Nash Equilibrium points (NE) of the stage game [12]. When there are multiple NEs in a game and the agents (players) are smart enough, the result of the game will be the unique Pareto optimal NE (PNE), if it exists [13].

 Therefore, we propose a scheme, to use independent QL agents for NOMA power allocation in an anti-jamming two-cell NOMA network.

\subsection{Related works}
Power allocation has an influential role in NOMA networks.
In [1] the optimal power allocation in Multi-Input Multi-Output
(MIMO) NOMA system, with statistical Channel State
Information at the Transmitter (CSIT), has been studied. The
authors of [2], have proposed a MIMO-NOMA scheme with
interference alignment. In this scheme, a low complexity
power allocation algorithm is used in each group of users.

Game theory studies can be useful to design
communicational schemes in anti-jamming communication
and multi-cell scenarios and many other frameworks in
wireless communication [10], $[14]-[18]$. For instance, in [18]
the NOMA power allocation problem in a heterogeneous
network is modeled as a sequential game between the Macro
BSs (MBS) and
Small BSs (SBS). And in [15] the maximization of a wireless
network's capacity is studied from a game theory perspective.

Using RL based algorithms in communicational systems,
has attracted a lot of attention, due to the ability to solve
extremely sophisticated problems, without the need for any
data sets. For instance, the works in [6], [9], $[19]-[22]$. In [6],
a fast QL based power allocation algorithm is designed to
solve the anti-jamming NOMA power allocation problem in a
MIMO-NOMA network. authors of [9], have proposed a DQL
based two-dimensional communication scheme, for an anti-jamming communication scenario.

\subsection{Novelties and Contributions}

The contributions of this paper can be summarized as follows :

\begin{itemize}[\IEEEsetlabelwidth{5}]
\item [$\bullet$]  We model the game between two BSs and a smart jammer in an anti-jamming two-cell NOMA network and drive the solutions of the game.

\item [$\bullet$]  We derive the unique Pareto NE point of the BSs game.

\item [$\bullet$]  A QL based unselfish (QLU) NOMA power allocation scheme, is proposed to be used in the two- cell anti-jamming NOMA scenario.

\item [$\bullet$]  We propose a Deep QL based Unselfish (DQLU) NOMA power allocation scheme to be used in the two-cell anti-jamming NOMA scenario.

\item [$\bullet$]
We also develop DQLU by using the hot booting technique as HBDQLU scheme.
\end{itemize}

The remainder of this paper is organized as follows: The system model is presented in
section \ref{System model}. The sequential game between two BSs and the smart
jammer, in a two-cell NOMA network is modeled in section
\ref{The sequential game}, also the solutions of this game, are derived in this section.
QLU, DQLU HBDQLU scheme for NOMA power allocation are proposed in
section \ref{NOMA POWER ALLOCATION}. Section \ref{Simulation results} discusses the simulation results. The
conclusions are presented in Section \ref{Conclusion}. The proofs for the
solutions of the game are provided in Appendices.



\section{System Model}\label{System model}
As shown in Fig. 1, we considered a NOMA network consisted
of two cells. Where the Base Station (BS) of each cell, serves
two Users (UE)s. BS1 serves UE1 and UE2, and BS2 serves
UE3 and UE4. There is inter-cell interference in this network.
The BSs are smart and can determine their total power and the
allocated power to each user. The task of power allocation is
done in each BS independently and the BSs, are not
coordinated with each other. The BSs just have  limited
communication with each other, to share the information of
the SINR of their users. Also, a smart jammer is considered in
this system model which, can choose its optimal power based
on the ongoing legal communication.

It is assumed that the BSs do not have instantaneous
channel state information. They do the power allocation and
successive interference cancelation ordering, based on the
SINR, which the users send on the feedback channels. Let $h_{i}^{1}, h_{i}^{2}$ and $h_{i}^{J}$, be the channels between user UEi and
$\mathrm{B}\mathrm{S}1$, BS2 and jammer respectively. the channels are Gaussian
distributed and follow $h_i^\mu  \sim G(0,L)$ , with $\mu=1,2, J$, where,
$L$ is the large scale fading factor and is given by (1).

\begin{equation}
\label{}
L = {\textstyle{{{{10}^{ - 3.53}}} \over {{{\left| {d_i^\mu } \right|}^{3.76}}}}},
\end{equation}
$d_{i}^{\mu}$: the distance between source $\mu$ and receiver UEi.
\begin{figure}
  \includegraphics[width=3.5in]{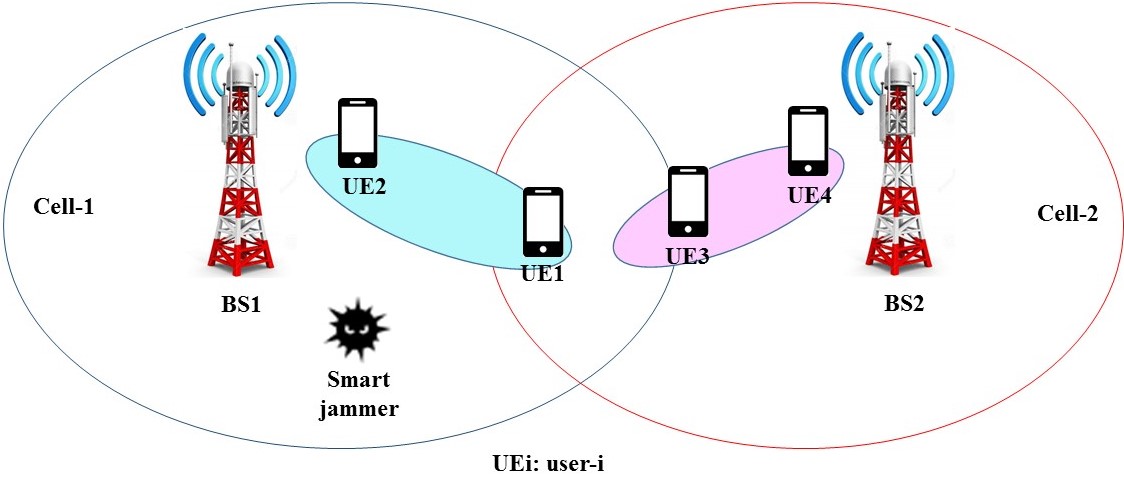}
  \caption{Downlink NOMA transmission against smart jammer, in a two-cell network.}\label{Systemmodel}
\end{figure}

Each BS sends the superimposed signals to all of its users. Each user at first, decodes the signal of weaker
users, according to the Successive Interference Cancelation
(SIC) decoding order. Then subtracts the components of the
decoded signals from the received signal. So this user can
decode its own signal without the interference of the signal of weaker users. More specifically, UE2 and UE4, at first,
decode the signals of UE1 and UE3 respectively. Then, UE2
and UE4 subtract the components of these signals, from the
received signal by each of them, and finally, they decode their
own signals. We assume that the inter-cell interference and
the jamming signal, do not change the optimal SIC decoding
order [6]. Thus, (2) and (3), must be held [6].
\begin{equation}
\label{}
\frac{{{{\left| {h_1^1} \right|}^2}}}{{1 + {p_{BS2}}{{\left| {h_1^2} \right|}^2} + {p_J}{{\left| {h_1^J} \right|}^2}}} \le \frac{{{{\left| {h_2^1} \right|}^2}}}{{1 + {p_{BS2}}{{\left| {h_2^2} \right|}^2} + {p_J}{{\left| {h_2^J} \right|}^2}}}
\end{equation}

where $p_{J}$ is the power of jammers signal and $p_{BS1}$ and $p_{BS2}$ are the total transmission power of BS1 and BS2, respectively.
\begin{equation}
\label{}
\frac{{{{\left| {h_3^2} \right|}^2}}}{{1 + {p_{BS1}}{{\left| {h_3^1} \right|}^2} + {p_J}{{\left| {h_3^J} \right|}^2}}} \le \frac{{{{\left| {h_4^2} \right|}^2}}}{{1 + {p_{BS1}}{{\left| {h_4^1} \right|}^2} + {p_J}{{\left| {h_4^J} \right|}^2}}}
\end{equation}

The smart jammer by sending a jamming signal, denoted by
$J$, aims to reduce the data sum rate of the total network. This
jammer can choose the optimal power $(p_{J})$, with the
maximum power constraint ${p}_{J}\leq P_{J}$. Considering the cost of
sending the jamming signal, and the goal of jamming, the
optimal jamming power can be chosen, based on the ongoing communication.

Let $x^{i}$ be the transmitted signal for UEi, and $x^{1}$ and $x^{2}$ be the
superimposed signal, transmitted from BS1 and BS2, respectively. $n_{i}$ is the additive Gaussian noise, ${n_i} \sim G(0,{\sigma ^2})$ , where, $\sigma^{2}$ is the variance of the noise. $y_{i}$
denotes the signal received by user UEi in  (4) to (7).
For UE2 and UE4, the receiving signals after SIC will be $y_{2}'$
and $y_{4}'$, respectively, and are given in (8) and (9).
\begin{equation}
\label{}
y_1= h_1^1{x^1} + h_1^2{x^2} + h_1^JJ + {n_1},
\end{equation}
\begin{equation}
\label{}
{y_2} = h_2^1{x^1} + h_2^JJ + {n_2},
\end{equation}
\begin{equation}
\label{}
{y_3} = h_3^1{x^1} + h_3^2{x^2} + h_3^JJ + {n_3},
\end{equation}
\begin{equation}
\label{}
{y_4} = h_4^2{x^2} + h_4^JJ + {n_4},
\end{equation}
\begin{equation}
\label{}
y_{2}' = h_2^1{x_2} + h_2^JJ + {n_2},
\end{equation}
\begin{equation}
\label{}
y_{4}' = h_4^2{x_4} + h_4^JJ + {n_4}.
\end{equation}

The data rate of the users is calculated according to  (5) and (6), where $R_{i}$ denotes the data rate of the UEi.
Then, we can write
\begin{equation}
\label{}
{R_i} = {\log _2}\left( {1 + SIN{R_i}} \right),
\end{equation}

\begin{equation}
\label{}
{R_1} = \log _2^{}\left( {1 + \frac{{{p_1}{{\left| {h_1^1} \right|}^2}}}{{1 + {p_2}{{\left| {h_1^1} \right|}^2} + ({p_3} + {p_4}){{\left| {h_1^2} \right|}^2} + {p_J}{{\left| {h_1^J} \right|}^2}}}} \right),
\end{equation}

\begin{equation}
\label{}
{R_2} = \log _2^{}\left( {1 + \frac{{{p_2}{{\left| {h_2^1} \right|}^2}}}{{1 + {p_J}{{\left| {h_2^J} \right|}^2}}}} \right),
\end{equation}

\begin{equation}
\label{}
{R_3} = \log _2^{}\left( {1 + \frac{{{p_3}{{\left| {h_3^2} \right|}^2}}}{{1 + {p_4}{{\left| {h_3^2} \right|}^2} + ({p_1} + {p_2}){{\left| {h_3^1} \right|}^2} + {p_J}{{\left| {h_3^J} \right|}^2}}}} \right),
\end{equation}

\begin{equation}
\label{}
{R_4} = \log _2^{}\left( {1 + \frac{{{p_4}{{\left| {h_4^2} \right|}^2}}}{{1 + {p_J}{{\left| {h_4^J} \right|}^2}}}} \right),
\end{equation}
where $\mathrm{S}\mathrm{I}\mathrm{N}\mathrm{R}_{\mathrm{i}}$ is the signal to noise plus interference ratio of the UEi and
the data rates are calculated in terms of (bit/s/Hz).
The main problem in this work is, to maximize the data sum
rate of the total network, under jamming attack and inter-cell
interference, and with the constraint of maximum total power
and the QoS.
\begin{subequations}
\begin{align}
(P1):& {\rm{  }}\mathop {\max }\limits_{{p_1},{p_2},{p_3},{p_4}} ({R_1} + {R_2} + {R_3} + {R_4}),\\
s.t.   {\rm{ (C1)\! :  }}&{\rm{ \; }} {p_{BS1}},{p_{BS2}} \le {p_{BS,\max }},\\
s.t.   {\rm{ (C2)\!: }} &{\rm{ \; }} \min ({R_1},{R_2},{R_3},{R_4}) \ge {R_0},
\end{align}
\end{subequations}
$R_{0}$ is the minimum data rate, demanded by the network. This
problem is equivalent to the following problem (P2).

\begin{subequations}
\begin{align}
\begin{array}{l}
(P2):{\rm{  }}\mathop {\max }\limits_{{p_1},{p_2},{p_3},{p_4}} (I\left( {\min ({R_1},{R_2},{R_3},{R_4}) \ge {R_0}} \right)\\
 \times  ({R_1} + {R_2} + {R_3} + {R_4})),
\end{array}\\
s.t.{\rm{ (C1)\!:  }}{\rm{ \; }}{p_{BS1}},{p_{BS2}} \le {p_{BS,\max }},
\end{align}
\end{subequations}
where I $\left( . \right)$ is the indicator function. If its inner criteria hold,
the amount of this function will be equal to 1. Otherwise, the
amount of this function will be $0$.

\section{Sequential Game Between Two BSs and a Smart Jammer}\label{The sequential game}

The interactions between two smart BSs and a smart
jammer can be modeled as a sequential game. In this game,
the BSs (as the leaders) choose their strategies at first, then the
smart jammer (as the follower) chooses its optimal strategy
according to the strategies of the BSs.

\begin{table}[!t]
\renewcommand{\arraystretch}{1.3}
\caption{Symbols and notation}
\label{table-one}
\centering
\begin{tabular}{|l|l|}
\hline
\multicolumn{1}{|l|}{a}&	\multicolumn{1}{|l|}{Strategy vector of each BS}	\\
\hline
\multicolumn{1}{|l|}{$h_{k}^{i}$}&	\multicolumn{1}{|l|}{Power gain of the channel between source-i and user-k}	\\
\hline
\multicolumn{1}{|l|}{$J$}&	\multicolumn{1}{|l|}{Signal of the smart jammer}	\\
\hline
\multicolumn{1}{|l|}{$n$}&	\multicolumn{1}{|l|}{Additive noise}	\\
\hline
\multicolumn{1}{|l|}{$P_{J}$}&	\multicolumn{1}{|l|}{Maximum jamming power}	\\
\hline
\multicolumn{1}{|l|}{$P_{k}$}&	\multicolumn{1}{|l|}{The power allocated to user-k}	\\
\hline
\multicolumn{1}{|l|}{$p_{BS}$}&	\multicolumn{1}{|l|}{Total transmission power of each BS}	\\
\hline
\multicolumn{1}{|l|}{$\mathrm{s}$}&	\multicolumn{1}{|l|}{State vector of each BS}	\\
\hline
\multicolumn{1}{|l|}{$U$}&	\multicolumn{1}{|l|}{Utility functions}	\\
\hline
\multicolumn{1}{|l|}{$\sigma^{2}$}&	\multicolumn{1}{|l|}{Power of the noise}	\\
\hline
\multicolumn{1}{|l|}{$6$}&	\multicolumn{1}{|l|}{The set of neural network weight}	\\
\hline
\multicolumn{1}{|l|}{$\epsilon$}&	\multicolumn{1}{|l|}{Exploration factor}	\\
\hline
\multicolumn{1}{|l|}{$\delta$}&	\multicolumn{1}{|l|}{Discount factor for future rewards}	\\
\hline
\multicolumn{1}{|l|}{$\gamma$}&	\multicolumn{1}{|l|}{The cost of sending jamming signal with base power}	\\
\hline
\multicolumn{1}{|l|}{$\Omega$}&	\multicolumn{1}{|l|}{The space of the strategies vector}	\\
\hline
\end{tabular}
\end{table}

\begin{equation}
G = \left\langle {\left\{ {B{S_1},B{S_2},J} \right\},\left\{ {{{\bf{a}}_1},{{\bf{a}}_2},{a_J}} \right\},\left\{ {{U_1},{U_2},{U_J}} \right\}} \right\rangle.
\end{equation}

It is considered that the BSs can communicate the
estimated SINR of their users to each other. But they make
decisions independently. Therefore they are considered as separate players
in the game. The utility functions of BSl and BS2 are denoted
by $U_{1}$ and $U_{2}$ respectively. These utility functions are given
by  (18). The utility function of the smart jammer
$(U_{J})$ is given by (19).
\begin{equation}\label{finv_dc_comp_5}
\begin{split}
{U_2} = {U_1} =& {I_z}\left( {\min ({R_1},{R_2}) > {R_0}} \right) \times {I_z}\left( {\min ({R_3},{R_4}) > {R_0}} \right)\\
&\times ({R_1} + {R_2} + {R_3} + {R_4} + \gamma {p_J}{\rm{)}},
\end{split}
\end{equation}
where  $I_z\left( . \right)$ is an indicator function. If it inner criteria holds,
the amount of this function will be equal to 1. Otherwise, the
amount of this function will be $z<<1$.
\begin{equation}
{U_J} =  - \left( {{R_1} + {R_2} + {R_3} + {R_4} + \gamma {p_J}} \right).
\end{equation}

Each BS can determine its total transmitted power and
also the allocated power to each user. The strategy vectors and related feasible sets for
BS1, BS2, and the smart jammer, are respectively given by
 (20), (21), and (22).
\begin{equation}
{{\bf{a}}_1} = {\left[ {{p_1}{\rm{   }}{p_2}} \right]^T}{\rm{ ,  }}{\rm{ \; }}{p_1},{p_2} > 0{\rm{,}}{\rm{ \; }}{p_{BS1}} = {p_1} + {p_2} \le {p_{BS,\max }},
\end{equation}
\begin{equation}
{{\bf{a}}_2} = {\left[ {{p_3}{\rm{   }}{p_4}} \right]^T}{\rm{ ,  }}{\rm{ \; }}{p_3},{p_4} > 0{\rm{,}}{\rm{ \; }}{p_{BS2}} = {p_3} + {p_4} \le {p_{BS,\max }},
\end{equation}
\begin{equation}
{a_J} = {p_J}{\rm{      ,}}{\rm{ \; }}0 \le {p_J} \le {P_J}.
\end{equation}

To find the NE of the whole game, at
first the optimal strategy of the smart jammer, must be derived
as a function of the strategies of the BSs. Then, by finding the
NE points of the leaders game (${NE}_L$), the solution is
completed. In any channel condition, the optimal strategy of
the smart jammer is $p_{J}'$ which is given by (23), (24). For proof see Appendix A.
\begin{align}
&\frac{{\partial {U_J}}}{{\partial {p_J}}}\left| {\begin{array}{*{20}{c}}
{}\\
{{p_J} = {p_J'}}
\end{array}} \right. = - \frac{{{{\left| {h_2^J} \right|}^2}}}{{{{\ln }^2}(1 + {p_J'}{{\left| {h_2^J} \right|}^2})}}\nonumber\\
 &- \frac{{{{\left| {h_1^J} \right|}^2}}}{{{{\ln }^2}(1 + {p_2}{{\left| {h_1^1} \right|}^2} + {p_{BS2}}{{\left| {h_1^2} \right|}^2} + {p_J'}{{\left| {h_1^J} \right|}^2})}}\nonumber\\
 &- \frac{{{{\left| {h_3^J} \right|}^2}}}{{{{\ln }^2}(1 + {p_4}{{\left| {h_3^2} \right|}^2} + {p_{BS1}}{{\left| {h_3^1} \right|}^2} + {p_J'}{{\left| {h_3^J} \right|}^2})}}\nonumber\\
 &- \frac{{{{\left| {h_4^J} \right|}^2}}}{{{{\ln }^2}(1 + {{p'}_J}{{\left| {h_4^J} \right|}^2})}} + \gamma  = 0,
\end{align}

\begin{equation}\label{diode_current_waveform_finv}
p_J^* =
\begin{cases}
{{P_J}} , & {{\rm{}}{p_J'} \ge P-J}\\
{{p_J'}{\rm{ }}}, & {{\rm{}}{P_J} \ge {{p'}_J} \ge 0}\\
0, & {{\rm{}}{p_J'} \le 0}
\end{cases}.
\end{equation}

Later, the NEs between BS1 and BS2, are derived in
different moods of the environment. In this derivation, the power
of the smart jammer according to  (24), is considered
as a function of strategies of the BSs.

Mood 1: in this mood, it is considered that there are some
pairs of $(p_{BS1,i},p_{BS2,i})$ such that  (25) to
(28), can hold simultaneously. The set of these pairs is
denoted by PS.
\begin{equation}
\log _2^{}\left( {1 + \frac{{({p_{BS1}} - {p_1}){{\left| {h_2^1} \right|}^2}}}{{1 + {p_J^*}{{\left| {h_2^J} \right|}^2}}}} \right) \ge {R_0},
\end{equation}
\begin{equation}
\begin{split}
&\log _2^{}\left( {1 + \frac{{{p_1}{{\left| {h_1^1} \right|}^2}}}{{1 + ({p_{BS1}} - {p_1}){{\left| {h_1^1} \right|}^2} + {p_{BS2}}{{\left| {h_1^2} \right|}^2} + {p_J^*}{{\left| {h_1^J} \right|}^2}}}} \right)\\
 &\ge {R_0},
\end{split}
\end{equation}
\begin{equation}
\log _2^{}\left( {1 + \frac{{({p_{BS2}} - {p_3}){{\left| {h_4^2} \right|}^2}}}{{1 + {p_J^*}{{\left| {h_4^J} \right|}^2}}}} \right) \ge {R_0},
\end{equation}
\begin{equation}
\begin{split}
&\log _2^{}\left( {1 + \frac{{{p_3}{{\left| {h_3^2} \right|}^2}}}{{1 + ({p_{BS2}} - {p_3}){{\left| {h_3^2} \right|}^2} + {p_{BS1}}{{\left| {h_3^1} \right|}^2} + {p_J^*}{{\left| {h_3^J} \right|}^2}}}} \right)\\
 &\ge {R_0}.
\end{split}
\end{equation}
Any strategy profile like $(\bf{a}_{1,i},\bf{a}_{2,i})$, where
$(p_{BS1,i},p_{BS2,i})\in PS$ and this strategy profile held in the following equations is an
${NE}_{L}$. For the proof, see Appendix B. This set of ${NE}_{L}$s is denoted by ${NE}_{L}$1.
\begin{equation}\label{finv_dc_comp_5}
\begin{split}
 & eq1 : ( {\frac{{{{\left| {h_2^1} \right|}^2}}}{{1 + {p_J^*}{{\left| {h_2^J} \right|}^2}}}}) \times ( {\frac{{{{\left| {h_4^2} \right|}^2}}}{{1 + {p_J^*}{{\left| {h_4^J} \right|}^2}}}}) \times {2^{(\gamma {p_J^*} + 2{R_0})}}  \\
 &\times( - \frac{{2{p_{BS1}}}}{{{2^{{R_0}}}}} \times \frac{{{{\left| {h_3^1} \right|}^2}}}{{{{\left| {h_3^2} \right|}^2}}} + (\frac{1}{{{2^{2{R_0}}}}} + \frac{{{{\left| {h_3^1} \right|}^2}}}{{{{\left| {h_3^2} \right|}^2}}} \times \frac{{{{\left| {h_1^2} \right|}^2}}}{{{{\left| {h_1^1} \right|}^2}}}) \times {p_{BS2}}\\
&- \frac{1}{{{2^{{R_0}}}}} \times \frac{{1 + {p_J^*}{{\left| {h_3^J} \right|}^2}}}{{{{\left| {h_3^2} \right|}^2}}} + \frac{{{{\left| {h_3^1} \right|}^2}}}{{{{\left| {h_3^2} \right|}^2}}} \times \frac{{1 + {p_J^*}{{\left| {h_1^J} \right|}^2}}}{{{{\left| {h_1^1} \right|}^2}}}),
\end{split}
\end{equation}
\begin{equation}\label{}
\begin{split}
& eq2 : (\frac{{{{\left| {h_2^1} \right|}^2}}}{{1 + {p_J^*}{{\left| {h_2^J} \right|}^2}}} \times \frac{{{{\left| {h_4^2} \right|}^2}}}{{1 + {p_J^*}{{\left| {h_4^J} \right|}^2}}})\times {2^{(\gamma {p_J^*} + 2{R_0})}} \\
 &\times( - \frac{{2{p_{BS2}}}}{{{2^{{R_0}}}}} \times \frac{{{{\left| {h_1^2} \right|}^2}}}{{{{\left| {h_1^1} \right|}^2}}} + (\frac{1}{{{2^{2{R_0}}}}} + \frac{{{{\left| {h_3^1} \right|}^2}}}{{{{\left| {h_3^2} \right|}^2}}} \times \frac{{{{\left| {h_1^2} \right|}^2}}}{{{{\left| {h_1^1} \right|}^2}}}) \times {p_{BS1}}\\
 &- \frac{1}{{{2^{{R_0}}}}} \times \frac{{1 + {p_J^*}{{\left| {h_1^J} \right|}^2}}}{{{{\left| {h_1^1} \right|}^2}}} + \frac{{{{\left| {h_1^2} \right|}^2}}}{{{{\left| {h_1^1} \right|}^2}}} \times \frac{{1 + {p_J^*}{{\left| {h_3^J} \right|}^2}}}{{{{\left| {h_3^2} \right|}^2}}})),
\end{split}
\end{equation}
\begin{equation}
\begin{split}
&\log _2^{}(1 + \frac{{{p_1}{{\left| {h_1^1} \right|}^2}}}{{1 + ({p_{BS1}} - {p_1}){{\left| {h_1^1} \right|}^2} + {p_{BS2}}{{\left| {h_1^2} \right|}^2} + {p_J^*}{{\left| {h_1^J} \right|}^2}}})\\
& = {R_0},
\end{split}
\end{equation}
\begin{equation}
\begin{split}
&\log _2^{}(1 + \frac{{{p_3}{{\left| {h_3^2} \right|}^2}}}{{1 + ({p_{BS2}} - {p_3}){{\left| {h_3^2} \right|}^2} + {p_{BS1}}{{\left| {h_3^1} \right|}^2} + {{p'}_J}{{\left| {h_3^J} \right|}^2}}}) \\
&= {R_0},
\end{split}
\end{equation}
\begin{equation}\label{}
\begin{split}
&((eq1 \ge 0) \wedge (eq2 \ge 0) \wedge ({R_2}(i) \ge {R_0}) \wedge ({R_4}(i) \ge {R_0}))\\
 &\vee ((eq1 < 0) \wedge (eq2 \ge 0) \wedge ({R_2}(i) = {R_0}) \wedge ({R_4}(i) \ge {R_0}))\\
 &\vee ((eq \ge 10) \wedge (eq2 < 0) \wedge ({R_2}(i) \ge {R_0}) \wedge ({R_4}(i) = {R_0}))\\
&\vee ((eq < 0) \wedge (eq2 < 0) \wedge ({R_2}(i) = {R_0}) \wedge ({R_4}(i) = {R_0}))\\
& = True.
\end{split}
\end{equation}

The strategy profile $(\overline{\bf{a}}_{1,i},\overline{\bf{a}}_{2,i})$ given by (34), is one of the ${NE}_{L}$ points. Because if each BS deviates from this strategy profile, while the other does not change its current strategy, the utility of  both of the BSs will
decrease. This strategy profile is the unique Pareto optimal
${NE}_{L}$ profile of the game in this mood. For proof see Appendix C. This strategy profile is
denoted by ${NE}_{L}$.
\begin{equation}
({{\bar p}_1},{{\bar p}_2},{{\bar p}_3},{{\bar p}_4}) = \mathop {\arg \max }\limits_{({p_1} + {p_2},{p_3} + {p_4}) \in PS} {U_1}.
\end{equation}

Mood 2: in this mood, it is considered that  (25) to
(28) cannot hold simultaneously. In other words, the data rate
of at least one of the users does not satisfy the QoS criteria of
the network. In this condition, any strategy that holds in
(35) to (38), is a ${NE}_{L}$. This set of equilibrium points is
denoted by ${NE}_{L}$2.
\begin{equation}
{p_3} + {p_4} = {p_{BS,\max }},
\end{equation}
\begin{equation}\label{}
\begin{split}
&\left( {\frac{1}{{{2^{{R_0}}}}}  \times {p_{BS1}} - \frac{{1 + {p_J^*}{{\left| {h_1^J} \right|}^2} + {p_{BS,\max }}{{\left| {h_1^2} \right|}^2}}}{{{{\left| {h_1^1} \right|}^2}}}} \right)\\
&\times \left( {\frac{{{{\left| {h_2^1} \right|}^2}}}{{1 + {p_J^*}{{\left| {h_2^J} \right|}^2}}}} \right) < {R_0},
\end{split}
\end{equation}
\begin{equation}\label{}
\begin{split}
&\left( {\frac{1}{{{2^{{R_0}}}}} \times {p_{BS,\max }} - \frac{{1 + {p_J^*}{{\left| {h_3^J} \right|}^2} + {p_{BS1}}{{\left| {h_3^1} \right|}^2}}}{{{{\left| {h_3^2} \right|}^2}}}} \right)\\
 &\times \left( {\frac{{{{\left| {h_4^2} \right|}^2}}}{{1 + {p_J^*}{{\left| {h_4^J} \right|}^2}}}} \right) \ge {R_0},
\end{split}
\end{equation}
\begin{equation}\label{}
\begin{split}
&\left( {\frac{1}{{{2^{{R_0}}}}} \times {p_{BS,\max }} - 2 \times \frac{{{{\left| {h_3^1} \right|}^2}}}{{{{\left| {h_3^2} \right|}^2}}} \times {p_{BS1}} - \frac{{1 + {p_J^*}{{\left| {h_3^J} \right|}^2}}}{{{{\left| {h_3^2} \right|}^2}}}} \right)\\
 &\times \left( {\frac{{{{\left| {h_4^2} \right|}^2}}}{{1 + {p_J^*}{{\left| {h_4^J} \right|}^2}}}} \right) \times \left( {\frac{{{{\left| {h_2^1} \right|}^2}}}{{1 + {p_J^*}{{\left| {h_2^J} \right|}^2}}}} \right) \ge 0.
\end{split}
\end{equation}

Consider $\overline{p}_{BS1}$ is the $p_{BS1}$ which makes (38) equal
to zero. In the set ${NE}_{L}$2, the ${NE}_{L}$ in which, $p_{BS1}$ has the
nearest amount to $\overline{p}_{BS1}$ is the Pareto optimal ${NE}_{L}$. This point is
denoted by ${PNE}_{L}$2.

Furthermore, any strategy profile that holds in
(39) to (42) is a NE. This set of equilibriums is denoted by
${NE}_{L}$3.
\begin{equation}
{p_1} + {p_2} = {p_{BS,\max }},
\end{equation}
\begin{equation}\label{}
\begin{split}
&\left( {\frac{1}{{{2^{{R_0}}}}} \times {p_{BS2}} - \frac{{1 + {p_J^*}{{\left| {h_3^J} \right|}^2} + {p_{BS,\max }}{{\left| {h_3^1} \right|}^2}}}{{{{\left| {h_3^2} \right|}^2}}}} \right)\\
 &\times \left( {\frac{{{{\left| {h_4^1} \right|}^2}}}{{1 + {p_J^*}{{\left| {h_4^J} \right|}^2}}}} \right) < {R_0},
\end{split}
\end{equation}
\begin{equation}\label{}
\begin{split}
&\left( {\frac{1}{{{2^{{R_0}}}}} \times {p_{BS,\max }} - \frac{{1 + {p_J^*}{{\left| {h_1^J} \right|}^2} + {p_{BS2}}{{\left| {h_1^2} \right|}^2}}}{{{{\left| {h_1^1} \right|}^2}}}} \right)\\
 &\times \left( {\frac{{{{\left| {h_2^1} \right|}^2}}}{{1 + {p_J^*}{{\left| {h_2^J} \right|}^2}}}} \right) \ge {R_0},
\end{split}
\end{equation}
\begin{equation}\label{}
\begin{split}
&\left( {\frac{1}{{{2^{{R_0}}}}} \times {p_{BS,\max }} - 2 \times \frac{{{{\left| {h_1^2} \right|}^2}}}{{{{\left| {h_1^1} \right|}^2}}} \times {p_{BS2}} - \frac{{1 + {p_J^*}{{\left| {h_1^J} \right|}^2}}}{{{{\left| {h_1^1} \right|}^2}}}} \right)\\
 &\times \left( {\frac{{{{\left| {h_4^2} \right|}^2}}}{{1 + {p_J^*}{{\left| {h_4^J} \right|}^2}}}} \right) \times \left( {\frac{{{{\left| {h_2^1} \right|}^2}}}{{1 + {p_J^*}{{\left| {h_2^J} \right|}^2}}}} \right) \ge 0.
\end{split}
\end{equation}

Consider $\overline{p}_{BS2}$ is the $p_{BS2}$ which makes (42) equal
to zero. In the set ${NE}_{L}$3, the ${NE}_{L}$ in which, $p_{BS1}$ has the
nearest amount to $\overline{p}_{BS2}$ is the Pareto optimal ${NE}_{L}$ this point is
denoted by ${PNE}_{L}$3. For the proof, see Appendix D.

\section{NOMA Power Allocation Schemes}\label{NOMA POWER ALLOCATION}
The BSs can use RL algorithms
to determine the total transmission power and the NOMA power
allocation strategy. The BSs can choose their strategies,
independently. Due to the cost of coordination, the need for a
central process unit, and the high dimensionality of the state-action spaces, the BSs are not coordinated. As mentioned
before, a smart jammer in each time slot has a determined
response to the strategies chosen by the BSs. Thus, a smart
jammer can be regarded as a part of a communicational
environment.

The repeated interactions between two or more independent
RL agents can be modeled as an evolutionary repeated game [12].
It is proven that in this kind of game the agents will
converge to the pure NEs of a stage game if they exist [12].
Furthermore, if the unique Pareto optimal NE
point exists, and the agents are smart enough, they will
converge to this point [13]. The existence of the unique ${PNE}_{L}$
point, in the game between BSs is proven in Section \ref{The sequential game}.

\subsection{QLU NOMA Power Allocation}
It is considered that the BSs do not have the instantaneous
CSI. Also, they do not have knowledge of the jamming
parameters. Therefore, they can use the Q-learning as a model-free
RL algorithm for power allocation. A lot of studies
have been done about QL, in the literature [6],[9]. Any QL based
algorithm can be specified by the state-action space, the
reward function, and the method of updating the Q-function.

The action sets of BSl and BS2 are given by  (20)
and (21), where the powers are quantized to $L_{p}$ levels.
When the QL agents choose their actions, the BSs transmit
the superimposed signals to their users simultaneously. Each
user after receiving its signal, estimates the SINR of the signal,
quantizes it, and sends it as feedback, to the relevant BS. In
this work, it is considered that the BSs have limited
communication with each other to share the SINR of their
users. Thus, the state vectors for the QL agents of BS1 and
BS2, are defined as $S_{1}^{k}=[SlNR_{1}^{k},SINR_{2}^{k},SINR_{3}^{k},SINR_{4}^{k}]$ and
$S_{2}^{k}=[S1NR_{3}^{k},S1NR_{4}^{k},S1NR_{1}^{k},S1NR_{2}^{k}]$ respectively. Where,
$SINR_{i}^{k}$ is the SINR of UEi in the kth time slot. The reward
functions of the QL-agents of BS1 and BS2, are equivalent to
the unselfish utility functions in (18).

Both of the QL agents, update their Q-functions,
according to  (43), and choose their actions with the
probability function in (44).

\begin{equation}\label{}
\begin{split}
&Q({{\bf{s}}^k},{{\bf{a}}^k}) \leftarrow {\rm{ }}(1 - \alpha )Q({{\bf{s}}^k},{{\bf{a}}^k}) + \alpha (U({{\bf{s}}^k},{{\bf{a}}^k})\\
 &+ \delta \mathop {\max }\limits_{{\bf{a}} \in \Omega } Q({{\bf{s}}^{k + 1}},{\bf{a}})),
\end{split}
\end{equation}
where $\alpha$ is the learning rate, $\delta$ is the discount factor, and $\Omega$ is
the space of action vectors.
\begin{equation}\label{}
pr({\bf{a}} = {\bf{\hat a}}) =
\begin{cases}
{1 - \varepsilon } , & {{\bf{\hat a}} = \arg \mathop {\max }\limits_{{\bf{a}} \in \Omega } Q({{\bf{s}}^k},{\bf{a}})}\\
\frac{\varepsilon }{{\left| \Omega  \right| - 1}}, & {\rm{ O}}{\rm{.W}}{\rm{.}}
\end{cases}.
\end{equation}

\subsection{DQLU NOMA Power Allocation}
When the state-action space is large as in this work, we
can use a DQL algorithm, to find the optimal policy [9]. In DQL
algorithms, two deep neural networks are used. One
of them is the main network and the other is the target
network. The target network is used in the training process of
the main network. And the main network is used to
approximate the Q-values of all of the actions in a state. Using
DQL algorithms in problems with a large state-action space
leads to better results than QL [9]. Since in DQL algorithms, the main neural network estimates the value of all actions in a particular state, in every single step, While the weights of the neural network get updated, by the interactions of the DQL-agent in the environment. But in QL algorithms in every step the Q-function of only a pair of state and action, get updated.

All of the considerations about using QL for power
allocation in the BSs and the interactions of the QL agents in
the communicational environment are true for DQL, apart
from (43). Instead, the estimation of Q-values
is the output of the main network which its weights and
biases are updating to minimize the loss given by  (45).
\begin{equation}
loss{\rm{ = targe}}{{\rm{t}}_{DQN}} - Q({\bf{s}},{\bf{a}}),
\end{equation}
\begin{equation}
{\rm{targe}}{{\rm{t}}_{DQN}} = U\left( {{\bf{s}},{\bf{a}}} \right) + \delta \mathop {\max }\limits_{{\bf{a'}}} T({\bf{s'}},{\bf{a'}}),
\end{equation}
where $T(\mathrm{s},\mathrm{a})$ is the Q-function approximated by the
target network. $\mathrm{s}'$ is the next state and $\mathrm{a}'$ is the action in the
next step.

The details of the main neural network used in the DQL
agents are shown in Fig. 2. As shown in this figure, the neural
network used in the DQLU algorithm consists of 4 layers.
The first layer is the input layer. It has as many nodes as the
size of the state vector. The second and third layers are hidden
layers, with 24 nodes in each one. The fourth layer is the output layer. There are as many nodes as $|\Omega|$. the activation
function of each layer is characterized by $\mathrm{R}$ as
ReLU and $\mathrm{L}$ as Linear.

\subsection{HBDQLU NOMA Power Allocation}
In the hot booting technique, the process of the DNN in DQL is initialized by the weights of a DNN which has worked in some similar scenarios. By using the hot booting technique, the DQL will have faster convergence and will get
better results, especially at the beginning of the process[9]. In this work, the DQL based, unselfish power allocation algorithm, which is using the hot booting technique, is called HBDQLU.

When the number of similar experiments for preparing the DNN in the hot booting technique increases, the resultant convergence of the system, increases. On the other hand, it will increase the risk of overfitting in some specific experiments [6]. Therefore, for choosing the number of scenarios for the preparing phase of the hot booting algorithm, the trade-off between the risk of over fitting and the resultant convergence rate, must be considered [6].
\begin{figure}
  \begin{center}
  \includegraphics[width=3in]{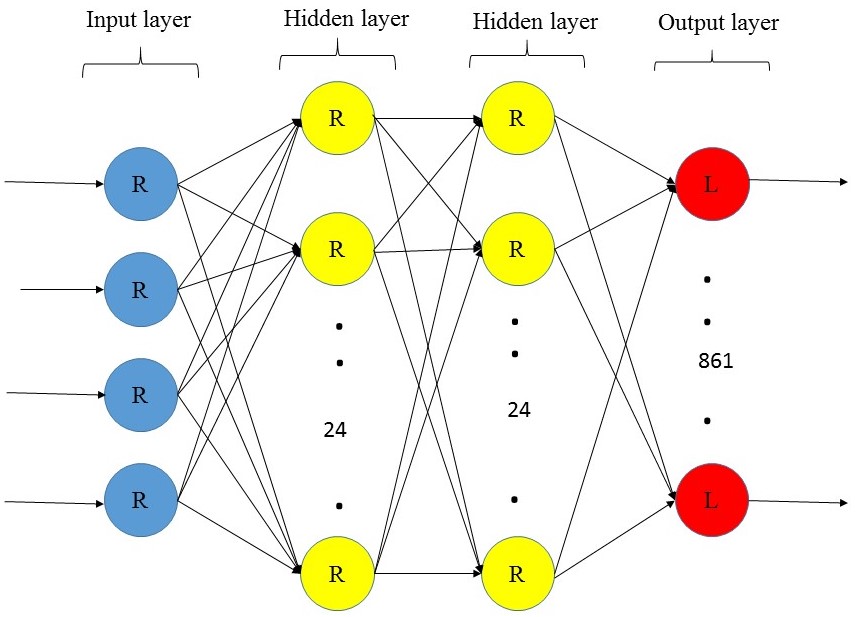}
  \caption{The Deep neural network used in DQLU scheme.}\label{DNN}
  \end{center}
\end{figure}

\section{Simulation results}\label{Simulation results}
In this section the performance of the purposed unselfish
NOMA power allocation, in the presence of a smart jammer, is
evaluated, using computer simulations. The QL based selfish
NOMA power allocation scheme is considered a benchmark.
We set $xl_{1}=250m, xl_{2}=20m, xl_{3}=400m$ , $xl_{4}=480m,$
$xl_{BS1}=0, xl_{BS2}=500m, \gamma=0.5, \delta=0.7, \sigma_{n}^{2}=-140db.$
Also we set $\alpha=0.2$ for QLU and $\alpha=0.1$ for DQLU and
HBDQLU. Where $xl_{j}$ denotes the location of the UEi. The
smart jammer is using a QL-based algorithm to determine its
jamming power level.

The gained rewards by BS1 and BS2, in DQLU, HBDQLU, and QLU schemes are shown in Fig. 3. The achieved rewards
by BS1 and BS2 in the QLS method, are shown in Fig. 4 and Fig. 5
respectively. Since the rewarding method is different for
unselfish and selfish methods, comparing them from this point
of view, is not correct. However, as shown in Fig. 3, DQLU
exceeds QLU, with 100\% higher reward, and HBDQLU
outperforms DQLU with an 18\% higher reward. As shown in Fig.
5, the reward of BS2 using QLS is increasing with time, until convergence. The reward of BS1, shown in Fig. 4, using the QLS method, is increasing with time at first, but after the 200th time slot, it gets decreasing with time slot number. Because BS2 in this method learns that by increasing its total transmit power, without considering the situation of the other BS, it can gain more reward. Thus, by increasing the total power of BS2, the condition of BS1 gets worse in such a way that, BS1 is not able to resist against jammer.

As shown in Fig. 7, QLU outperforms QLS with a 75\%  higher amount of objective function. HBDQLU and DQLU, improve the QLU sequentially with 100\% and 128\% higher amount of objective function.

The data sum-rate gained by the methods is shown in Fig. 6. As can be seen, DQLU and HBDQLU excide QLS, with 15\% and 19\%, higher data sum-rate. But the data sum-rate achieved by QLU is less than the data sum-rate achieved by QLS. It happens because, meeting QoS criteria, for the data rate of all of the users, has an intense impact on the reward of QLU. Thus in this method, the BSs try to find the strategies, which meet QoS for all of the users. But the QLU agents are not always able to find the strategies which, are admissible and optimal for data sum-rate, simultaneously.

\begin{figure}[!t]
  \centering
  \includegraphics[width=3in]{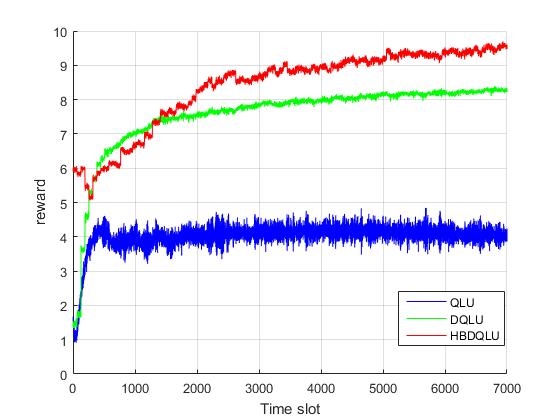}
  \caption{The gained reward by BS1 (and BS2) using in QLU, DQLU and HBDQLU, schemes.}\label{RewardBSsunselfish}
\end{figure}
\begin{figure}[!t]
  \centering
  \includegraphics[width=3in]{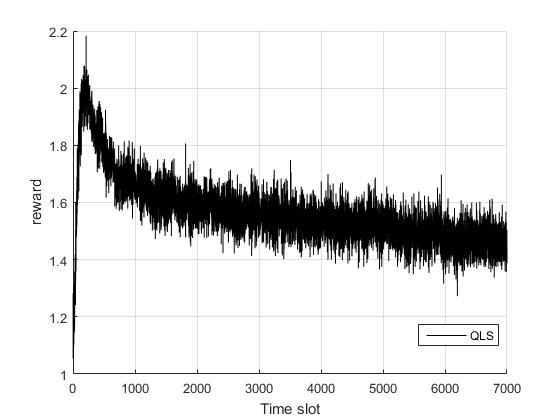}
  \caption{The gained reward by BS1, in QLS scheme.}\label{RewardBS1selfish}
\end{figure}
\begin{figure}[!t]
  \centering
  \includegraphics[width=3in]{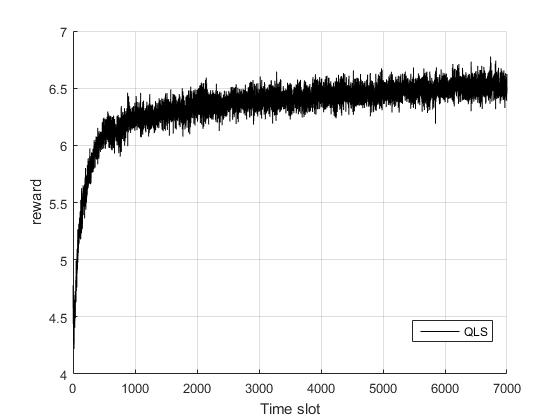}
  \caption{The gained reward by BS2, in QLS scheme.}\label{RewardBS2selfish}
\end{figure}

\begin{figure}[!t]
  \centering
  \includegraphics[width=3in]{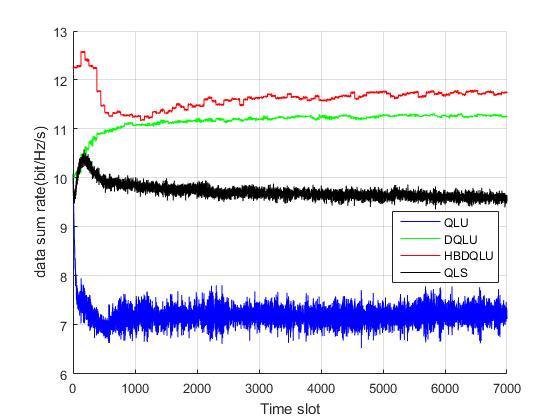}
  \caption{Data sum rate of the total network, in all of the methods.}\label{Datasumrate}
\end{figure}

\begin{figure}[!t]
  \centering
  \includegraphics[width=3in]{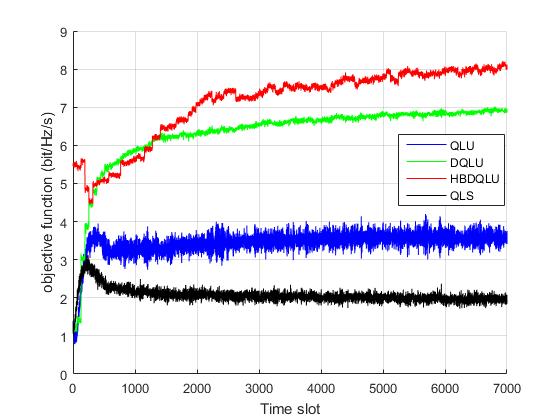}
  \caption{The amount of objective function of the total network.}\label{Objectivefunction}
\end{figure}

\section{Conclusion}\label{Conclusion}
The sequential game between two independent BSs (as the leaders) and a smart jammer (as the follower) has been modeled. The solution of this game has been derived in different conditions to prove the convergence of the proposed schemes. We proposed three new NOMA power allocation schemes to perform in the anti-jamming two-cell NOMA network. The schemes were QLU, DQLU, and HBDQLU. Using simulation results, tt was shown that the convergence of all of these schemes occurs after about 1000 time slots. We also showed that QLU outperforms QLS with a 75\% higher amount of objective function, and HBDQLU and DQLU improve the QLU sequentially with 100\% and 128\% higher amount of objective function, respectively.
As a future direction, we propose to design a constraint DQL or double DQL-based NOMA power allocation scheme. These methods might improve the convergence speed and also the final results gained by the proposed schemes. As another direction for future research, it can be proposed to model the game by considering more than one smart jammer in the network.

\appendices
\section{}
Equation (47) shows that the utility function of the jammer
is concave in terms of $p_{J}$ therefore, this function is
maximized when $p_{J}=p_{J}'$ and (23) holds. Since we have
$p_{J}\in[0,P_{J}]$, the optimal power of the jammer will be chosen
according to  (24).
\begin{align}
\frac{{{\partial ^2}{U_J}}}{{\partial p_J^2}} = &- \frac{{{{\left| {h_1^J} \right|}^4}}}{{{{\ln }^2}{{(1 + {p_2}{{\left| {h_1^1} \right|}^2} + {p_{BS2}}{{\left| {h_1^2} \right|}^2} + {p_J}{{\left| {h_1^J} \right|}^2})}^2}}}\nonumber \\
 &- \frac{{{{\left| {h_3^J} \right|}^4}}}{{{{\ln }^2}{{(1 + {p_4}{{\left| {h_3^2} \right|}^2} + {p_{BS1}}{{\left| {h_3^1} \right|}^2} + {p_J}{{\left| {h_3^J} \right|}^2})}^2}}}\nonumber\\
&- \frac{{{{\left| {h_2^J} \right|}^4}}}{{{{\ln }^2}(1 + {p_J}{{\left| {h_2^J} \right|}^2})}}\nonumber\\
& - \frac{{{{\left| {h_4^J} \right|}^4}}}{{{{\ln }^2}{{(1 + {p_J}{{\left| {h_4^J} \right|}^2})}^2}}}< 0.
\end{align}

\section{}
Here, it is proven that when the strategy profiles from ${NE}_L$1
are chosen by the BSs, none of them will have the incentive to
deviate from that strategy, assuming the other BS doesn't
change its strategy. This proof is done just for when we have
(48). The proof related to the other terms in  (33), can
be done in a similar way. According to (49) and
(50), we have $\displaystyle \frac{\partial U_{1}}{\partial p_{1}}>0$, $\displaystyle \frac{\partial U_{2}}{\partial p_{3}}>0$. It means that U1 and U2 are
monotonically decreasing with $p_{1}$ and $p_{3}$, respectively. Thus,
they do not have the incentive to increase $p_{1}$ and $p_{3}$ from the
lowest acceptable amount, which meets the  QoS criteria. The
allocated power to all of the users, as functions of $p_{BS1}$ and
$p_{BS2}$ are given in  (31) and (32).  Also in this way, U1
and U2 will be functions of $p_{BS1}$ and $p_{BS2}$. The rest of this
proof is done, by considering $\hat{U}_{1}=2^{U_{1}}$ and $\hat{U}_{2}=2^{U_{2}}$ instead
of $U_{1}$ and $U_{2}$ respectively. Since $\hat{U}_{1}$ and $\hat{U}_{2}$ increasing
monotonically with $U_{1}$ and $U_{2}$, this consideration is true in
this proof. Equations (51) and (52) show that $\hat{U}_{1}$ and $\hat{U}_{2}$ are
concave in terms of $p_{BS1}$ and, $p_{BS2}$. Therefore by considering
$eq1\geq 0 (\displaystyle \frac{\partial\hat{U}_{1}}{\partial p_{BS1}}\geq 0), eq2\geq 0 (\displaystyle \frac{\partial\hat{U}_{2}}{\partial p_{BS2}}\geq 0)$, each of the BSs does not
have the incentive to decrease its total power. On the other
hand, BS1 and BS2, will not have the motivation to increase
their total power, while the other BS, does not change its
strategy. Because it will lead to $R_{1}$ or $R_{3}$, to be less than $R_{0}.$
\begin{align}
&(eq1 \ge 0) \wedge (eq2 \ge 0) \wedge ({R_2}(i) \ge {R_0}) \wedge ({R_4}(i) \ge {R_0})\nonumber\\
&= True,
\end{align}
\begin{align}
&\frac{{\partial {U_1}}}{{\partial {p_1}}} = ({p_{BS1}}{\left| {h_1^1} \right|^2} - {p_{BS1}}{\left| {h_2^1} \right|^2}\nonumber \\
 & + {p_{BS1}}{\left| {h_1^1} \right|^2}(p_J^*{\left| {h_2^J} \right|^2} + {p_{BS2}}{\left| {h_2^2} \right|^2})\nonumber\\
& - {p_{BS1}}{\left| {h_2^1} \right|^2}(p_J^*{\left| {h_1^J} \right|^2} + {p_{BS2}}{\left| {h_1^2} \right|^2}))\nonumber\\
& \times ((1 + p_J^*{\left| {h_2^J} \right|^2} + {p_{BS2}}{\left| {h_2^2} \right|^2} + {p_{BS1}}{\left| {h_2^1} \right|^2}({p_{BS1}} - {p_1}))\nonumber\\
& \times (1 + p_J^*{\left| {h_1^J} \right|^2} + {p_{BS2}}{\left| {h_1^2} \right|^2} + {p_{BS1}}{\left| {h_1^1} \right|^2}({p_{BS1}} - {p_1})){)^{ - 1}} \nonumber\\
&< 0,
\end{align}
\begin{align}
&\frac{{\partial {U_2}}}{{\partial {p_3}}} = ({p_{BS2}}{\left| {h_3^2} \right|^2} - {p_{BS2}}{\left| {h_4^2} \right|^2}\nonumber \\
 & + {p_{BS2}}{\left| {h_3^2} \right|^2}(p_J^*{\left| {h_4^J} \right|^2} + {p_{BS1}}{\left| {h_4^1} \right|^2})\nonumber\\
& \times ((1 + p_J^*{\left| {h_4^J} \right|^2} + {p_{BS1}}{\left| {h_4^1} \right|^2} + {p_{BS2}}{\left| {h_4^2} \right|^2}({p_{BS2}} - {p_3}))\nonumber\\
& \times ((1 + p_J^*{\left| {h_2^J} \right|^2} + {p_{BS2}}{\left| {h_2^2} \right|^2} + {p_{BS1}}{\left| {h_2^1} \right|^2}({p_{BS1}} - {p_1}))\nonumber\\
&  \times {(1 + p_J^*{\left| {h_3^J} \right|^2} + {p_{BS1}}{\left| {h_3^1} \right|^2} + {p_{BS2}}{\left| {h_3^4} \right|^2}({p_{BS2}} - {p_3}))^{ - 1}}\nonumber\\
&< 0,
\end{align}
\begin{align}
\frac{{{\partial ^2}{{\hat U}_2}}}{{\partial p_{BS2}^2}} =&  - 2\frac{1}{{{2^{{R_0}}}}} \times \frac{{{{\left| {h_1^2} \right|}^2}}}{{{{\left| {h_1^1} \right|}^2}}} \times \left( {\frac{{{{\left| {h_2^1} \right|}^2}}}{{1 + {p_J^*}{{\left| {h_2^J} \right|}^2}}}} \right)\nonumber \\
& \times \left( {\frac{{{{\left| {h_4^2} \right|}^2}}}{{1 + {p_J^*}{{\left| {h_4^J} \right|}^2}}}} \right) \times {2^{(\gamma {p_J^*} + 2{R_0})}} < 0,
\end{align}
\begin{align}
\frac{{{\partial ^2}{{\hat U}_1}}}{{\partial p_{BS1}^2}} =&  - 2\frac{1}{{{2^{{R_0}}}}} \times \frac{{{{\left| {h_3^1} \right|}^2}}}{{{{\left| {h_3^2} \right|}^2}}} \times \left( {\frac{{{{\left| {h_2^1} \right|}^2}}}{{1 + {p_J^*}{{\left| {h_2^J} \right|}^2}}}} \right)\nonumber \\
& \times \left( {\frac{{{{\left| {h_4^2} \right|}^2}}}{{1 + {p_J^*}{{\left| {h_4^J} \right|}^2}}}} \right) \times {2^{(\gamma {p_J^*} + 2{R_0})}} < 0.
\end{align}

\section{}
According to  (49) and (50), $\overline{p}_{1}$ and $\overline{p}_{3}$ must be
given by  (31) and (32) respectively, for when,
$p_{BS1}=\overline{p}_{BS1}$ and $p_{BS2}=\overline{p}_{BS2}$. Therefore the maximization problem
in (34) can be represented as (53). PS1 is the set of (
$p_{BS1},p_{BS2})$ which hold in  (54) and (55). Since $\hat{U}_{1}$
and $\hat{U}_{2}$ are concave in terms of $p_{BS1}$ and $p_{BS2}$, while the
space of PSl is convex, (53) will have just one answer.
\begin{align}
({\bar p_{BS1}},{\bar p_{BS2}}) = \mathop {\arg \max }\limits_{({p_{BS1}},{p_{BS2}}) \in {P_1}} {\hat U_1},
\end{align}
\begin{align}
&\left( {\frac{1}{{{2^{{R_0}}}}} \times {p_{BS1}} - \frac{{1 + {p_J^*}{{\left| {h_1^J} \right|}^2} + {p_{BS2}}{{\left| {h_1^2} \right|}^2}}}{{{{\left| {h_1^1} \right|}^2}}}} \right)\nonumber \\
 &\times \left( {\frac{{{{\left| {h_2^1} \right|}^2}}}{{1 + {p_J^*}{{\left| {h_2^J} \right|}^2}}}} \right) \ge {R_0},
\end{align}
\begin{align}
&\left( {\frac{1}{{{2^{{R_0}}}}} \times {p_{BS2}} - \frac{{1 + {p_J^*}{{\left| {h_3^J} \right|}^2} + {p_{BS1}}{{\left| {h_3^1} \right|}^2}}}{{{{\left| {h_3^2} \right|}^2}}}} \right)\nonumber \\
 &\times \left( {\frac{{{{\left| {h_4^2} \right|}^2}}}{{1 + {p_J^*}{{\left| {h_4^J} \right|}^2}}}} \right) \ge {R_0}.
\end{align}

\section{}
This proof is done just for ${NE}_{L}$2. The proof for ${NE}_{L}$3 will
be done in the same manner. It is considered in mood2, that it is
not possible to satisfy QoS for all of the users. ${NE}_{L}$2 is the
solution of the game when we have $R_{1}<R_{0}$ and $R_{3}\geq R_{0}$. In
this mood, it can be proved (in the same manner as Appendix
B that $U_{1}$ and $U_{2}$, are monotonically decreasing with $p_{1}$ and
$p_{3}$ thus in any NE point, $p_{1}$ and $p_{3}$, must be
set at the lowest admissible amount. Then we have $p_{1}=0,$
and $p_{3}$ will be equal to the lowest amount, which makes $R_{3}$
equal to $R_{0}$ (equation (31)) consequently $\hat{U}_{1}=2^{U_{1}},\hat{U}_{2}=2^{U_{2}}$ ,
are given by (56). According to (57), $\hat{U}_{2}$ is
monotonically increasing with $p_{BS2}$ thus, BS2 will never have
the incentive to decrease its total power, from $p_{BS,\max}$. And
according to (58), $\hat{U}_{1}$ is concave in terms of $p_{BS1}$. Therefore
(38) shows that $\displaystyle \frac{\partial\hat{U}_{1}}{\partial p_{BS1}}\geq 0$. Thus, there is no reason for BS1 to decrease $p_{BS1}.\mathrm{O}\mathrm{n}$ the other hand, it will not
have the incentive to increase $p_{BS1}$ because, increasing it, will
lead $R_{3}$ to be less than $R_{0}.$
\begin{align}
{{\hat U}_1} = {{\hat U}_2} =& \left( {\frac{1}{{{2^{{R_0}}}}} \times {p_{BS2}} - \frac{{1 + {p_J^*}{{\left| {h_3^J} \right|}^2} + {p_{BS1}}{{\left| {h_3^1} \right|}^2}}}{{{{\left| {h_3^2} \right|}^2}}}} \right)\nonumber \\
 & \times \left( {\frac{{{{\left| {h_4^2} \right|}^2}}}{{1 + {p_J^*}{{\left| {h_4^J} \right|}^2}}}} \right) \times {2^{(\gamma {p_J^*} + {R_0} + {z_1})}},
\end{align}

\begin{align}
\frac{{\partial {{\hat U}_2}}}{{\partial {p_{BS2}}}} =& \left( {\frac{{{p_{BS1}}{{\left| {h_2^1} \right|}^2}}}{{1 + {p_J^*}{{\left| {h_2^J} \right|}^2}}}} \right) \times \left( {\frac{{{p_{BS2}}{{\left| {h_4^2} \right|}^2}}}{{1 + {p_J^*}{{\left| {h_4^J} \right|}^2}}}} \right)\nonumber\\
 &\times {2^{(\gamma {p_J^*} + {z_0})}} > 0,
\end{align}

\begin{align}
\frac{{{\partial ^2}{{\hat U}_1}}}{{\partial p_{BS1}^2}} =&  - \frac{{{{\left| {h_3^1} \right|}^2}}}{{{{\left| {h_3^2} \right|}^2}}} \times \left( {\frac{{{{\left| {h_2^1} \right|}^2}}}{{1 + {p_J^*}{{\left| {h_2^J} \right|}^2}}}} \right) \times \left( {\frac{{{{\left| {h_4^2} \right|}^2}}}{{1 + {p_J^*}{{\left| {h_4^J} \right|}^2}}}} \right)\nonumber\\
 & \times {2^{(1 + \gamma {p_J^*} + {z_1})}} < 0.
\end{align}



\ifCLASSOPTIONcaptionsoff
  \newpage
\fi



\end{document}